\documentclass{article}
\usepackage{spconf,amsmath,graphicx}
\usepackage{xcolor}

\usepackage{enumitem}
\setlist{nosep, leftmargin=14pt}
\usepackage{comment}
\usepackage{mwe} % to get dummy images
\usepackage{float}

\newcommand{\SubItem}[1]{
    {\setlength\itemindent{11pt} \item[-] #1}
}

\title{AirSPEC: An IoT-empowered air quality monitoring system integrated with a machine learning framework to detect and predict defined air quality parameters }

\name{Nuwan Bandara$^1$, Sahan Hettiarachchi$^1$ and Prabhani Athukorala$^1$}
\address{$^1$Department of Electronic and Telecommunication Engineering, University of Moratuwa, Sri Lanka}
\begin{document}
%\ninept
%
\maketitle
\begin{abstract}
\textbf{
The air that surrounds us is the cardinal source of respiration of all life-forms. Therefore, it is undoubtedly vital to highlight that the balanced air quality is utmost important to the respiratory health of all living beings, environmental homeostasis, and even economical equilibrium. Nevertheless, a gradual deterioration of air quality has been observed in the last few decades, due to the continuous increment of polluted emissions from automobiles and industries into the atmosphere. Even though many people have scarcely acknowledged the depth of the problem, the persistent efforts of determined parties, including the World Health Organization, have consistently pushed the boundaries for a qualitatively better global air homeostasis, by facilitating technology-driven initiatives to timely detect and predict air quality in regional and global scales. However, the existing frameworks for air quality monitoring lack the capability of real-time responsiveness and flexible semantic distribution. In this paper, a novel Internet of Things framework is proposed which is easily implementable, semantically distributive, and empowered by a machine learning model. The proposed system is equipped with a NodeRED dashboard which processes, visualizes, and stores the primary sensor data that are acquired through a public air quality sensor network and further, the dashboard is integrated with a machine-learning model to obtain temporal and geo-spatial air quality predictions. ESP8266 NodeMCU is incorporated as a subscriber to the NodeRED dashboard via a message queuing telemetry transport broker to communicate quantitative air quality data or alarming emails to the end-users who access the system through the developed web and mobile applications. Therefore, the proposed system could become highly beneficial in empowering the public engagement in air quality in a data-driven manner through an unoppressive semantic framework.} 
\end{abstract}
\section{Introduction}
\label{sec:intro}
The quality of atmospheric conditions is vital in the context of overall life-form existence including humankind, animals, and plants \cite{N1,N2}. Nonetheless, over the past years, a subsequent deterioration of air quality has been noticed due to the increasing emissions of pollutants into the atmosphere from industries, automobiles, and burnt areas. Even though many people have scarcely recognized the depth of the problem \cite{N3,N4}, the world health organization (WHO) continuously emphasizes the abominable statistics of the issue: $90\%$ of the population breaths polluted air while leading seven million deaths per year \cite{N5,N6}. Further, poor air quality results in a deleterious negative impact on the ecosystem of the planet such as causing to accelerate the depletion of atmospheric ozone layer \cite{N7}.

Recent scientific studies convey \cite{N8,N9} the correlation between the on-going corona virus disease 2019 (COVID-19) pandemic and air pollution, highlighting that particulate matter (PM) could carry the RNA samples of the SARS-CoV-2 virus \cite{N9} or the increase of COVID-19 mortality rate is associated with the increment of the concentration of $PM_{2.5}$ in the air \cite{N8}. Therefore, it is evident to state that the real-time accurate monitoring of air quality is essential for the sustainable long-term health of the humankind and the natural ecosystem and in addition, it also assists in the prevailing combat against the COVID-19 pandemic.

\begin{figure*}[t!]
\centering
\begin{minipage}[b]{1.0\linewidth}
  \centerline{\includegraphics[width=15cm]{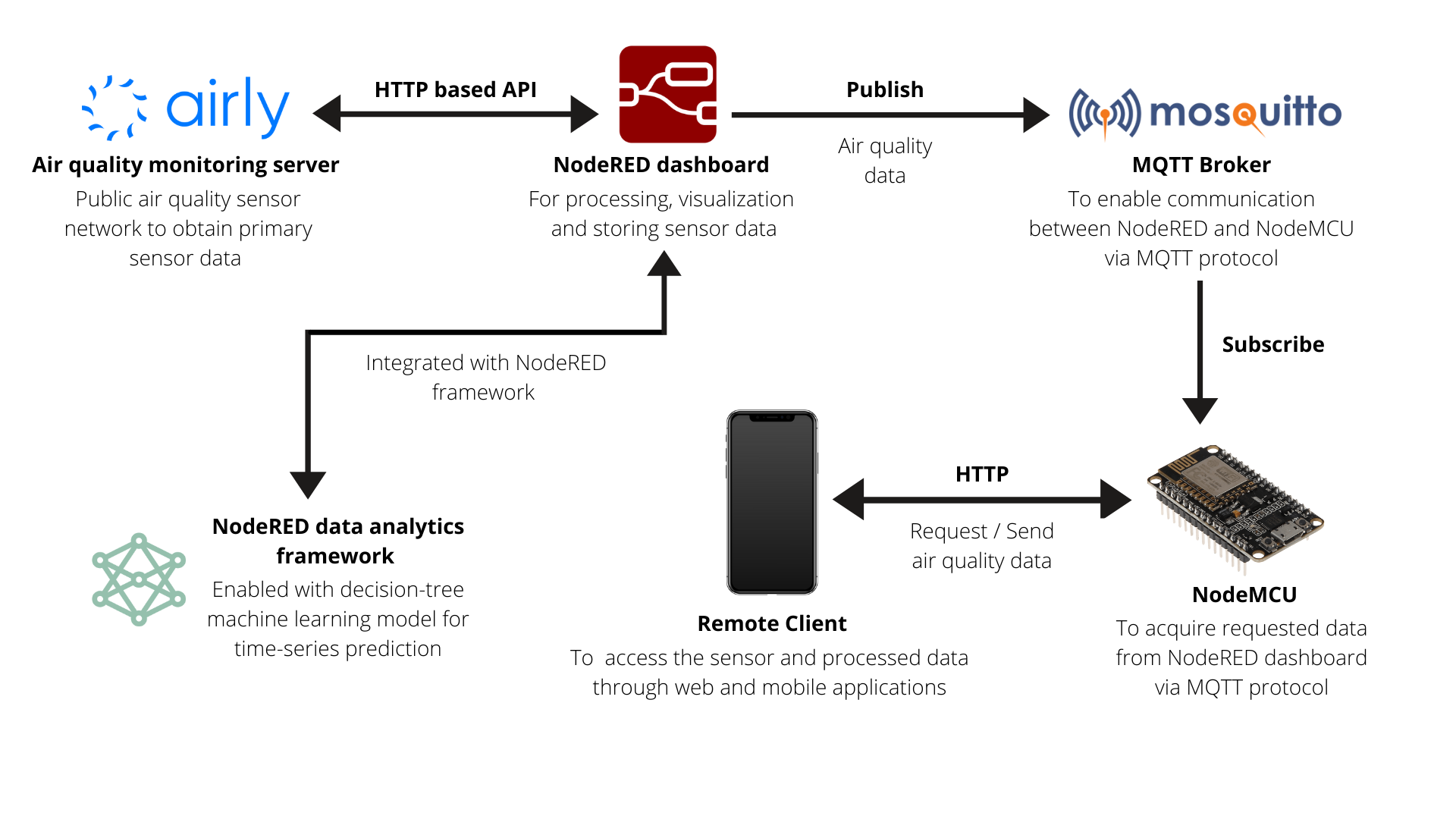}}
\end{minipage}
\vspace{-1.1cm}
\caption{The high-level architecture of the proposed system}
\label{fig:system}
\end{figure*}

Internet of Things (IoT) is a promising technology which facilitates the communication between objects, machines and people in a uniquely addressable manner via a set of standard communication protocols. Since IoT systems are dynamic, distributive, and built upon vast number of smart heterogeneous objects, the need for semantic inter-operability and energy optimization, while being scalable, is utmost important. The constrained environments of the IoT systems further stress the requirement of resource optimization in IoT systems, leading to light-weight communication protocols and low-power hardware implementations. These requirements further manipulate IoT as an integrative technology which could be implanted in various fields including real-time air quality monitoring. Even though many scholars have studied on utilizing IoT into air quality monitoring \cite{N10,N11,N12,N13,N14}, the semantic interpretation of the acquired data towards a futuristic perspective is yet to be explored in a thorough manner.

Thus, this paper presents a semantically distributive and easily implementable IoT predictive framework integrated with a machine learning model to detect and predict air quality parameters including $PM_1$, $PM_{2.5}$ and $PM_{10}$ along with temporal and spatial humidity, temperature and pressure distributions. The proposed system retrieves primary data through a public air quality sensor network, airly \cite{Nweb1}, and is equipped with a NodeRED dashboard \cite{Nweb2} which operates as a client of the sensor network to process, visualize and store the acquired air quality and weather data. The NodeRED dashboard is further responsible for delivering the predicative outputs via the embedded time-series decision-tree machine learning model in the dashboard back-end. The system is incorporated with a ESP8266 NodeMCU node \cite{Nweb3, Nweb4} to be operated as a subscriber to the NodeRED dashboard via message queuing telemetry transport (MQTT) protocol to deliver the quantitative air quality data to the end-users via publish-subscribe architecture. The end-users of the proposed system could access the sensor data as well as the predictions via quantitative and visualized formats via the developed mobile and web applications.       

\section{Methods}
\label{sec:pagestyle}

%%%%%%%%%%%%%%%%%%%%%%%%%%%%%%%%%%%%%%%%%%%%%%%%
\subsection{Background}
\subsubsection{Message Queuing Telemetry Transport Protocol}
Message Queuing Telemetry Transport \cite{don} is a lightweight publish-subscribe network protocol. For receiving messages that are published or sent under a particular topic by the publisher, the message receiver alias the subscriber should subscribe to the publisher under that specific topic. The main function of the MQTT is behaving as a broker that distributes received messages from the publisher to the specific subscriber by filtering the messages by topic. MQTT protocol can be applied in IoT applications under limited resources as it is a lightweight protocol. The Organization for the Advancement of Structured Information standard and ISO/IEC 20922 standard are recommended for the MQTT protocol.
\subsubsection{NodeRED framework}
NodeRED is a programming tool for wiring hardware devices, application programming interfaces (API) and online services together. It has a browser-based editor to manipulate a user-ergonomic platform to connect flows using a palette of nodes that can be deployed to the run-time with a single click. The users are enabled with stitching web services and hardware to each other by replacing common low-level coding tasks by NodeRED, through a visual drag-drop interface. Various components in NodeRED are connected to create such flows.

\subsubsection{NodeMCU module}
NodeMCU is an open-source based firmware and it is developed using ESP8266 which is a low cost Wi-Fi enabled chip. By exploring functionality on the ESP8266 chip, NodeMCU firmware comes with the ESP8266 development board. Since NodeMCU is an open-source platform, its hardware design is open for editing, modifying, and building.

\subsubsection{Decision-Tree based machine learning model}
Decision-tree learner is one of the widely utilized machine learning algorithms for classification purposes since of its ability to perform the task with less computational cost in both training and testing phases. In addition, the algorithm also guarantees the interpretability of the deduced models through the traditional recursive top-down induction of decision trees, in which the algorithm chooses the most effective data attribute to divide the dataset by considering the gain ratio criteria. The leaf of the tree is selected after reaching the pre-defined minimum number of instances through the dividing process. The generalization of the deduced model is performed thereafter in order to optimize the size of the model.

\subsection{System Overview}
At first, the airly air monitoring API is used to obtain the real-time sensor data of defined air quality parameters such as $PM_{1}$ and $PM_{2.5}$. The NodeRED framework is utilized for initial data processing and visualization obtained from airly API and it is supported by the deployed conventional decision-tree machine learning model which is implemented to predict the time-series values of the air parameters. Furthermore, the NodeRED framework communicates  with  the NodeMCU via the assigned MQTT broker and thus, with the mobile application in order to process the requests from the users within the implementation arena.

NodeRED framework is supported by the following special libraries which are related to different functions of the proposed system.

\begin{itemize}[noitemsep,nolistsep]
    \item   node-red-contrib-machine-learning-v2
    \SubItem{To build the machine learning model}
    \item   node-red-contrib-credentials
    \SubItem{To include credentials which are related to access airly}
    \item   node-red-node-email
    \SubItem{To send emergency emails where the air quality parameters of the requested location is above the corresponding safe levels}
    \item   node-red-contrib-fs
    \SubItem{To organize files which are related to obtaining primary sensor data and training the machine learning model}
    %\vspace{-0.2cm}
\end{itemize}

\begin{figure}[htb]
\begin{minipage}[b]{1.0\linewidth}
  \centering
  \centerline{\includegraphics[width=8.7cm]{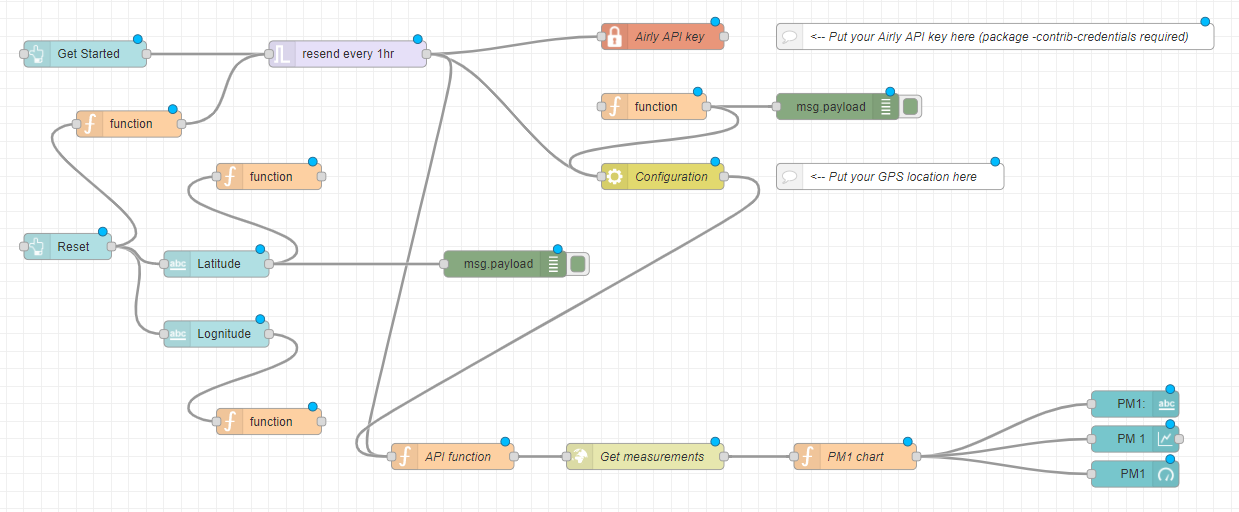}}
\end{minipage}
\caption{Flow management design for $PM_1$ in the NodeRED framework. The shown flow is integrated and extended to obtain and process the remaining air quality parameters.}
\label{fig:res}
\end{figure}

The users, who are able to access the NodeRED dashboard, can obtain primary sensor data for certain air quality parameters such as particulate matter, air quality index, temperature, pressure, and humidity based on location and time. In addition, users can view previous data as well as one-hour forecast data, which thus, simplifies any preferred decision-making process in relation to the user. The NodeRED dashboard also contains a file management system that allows users to see and download current and previous data files as needed.

\begin{figure}[htb]
\begin{minipage}[b]{1.0\linewidth}
  \centering
  \centerline{\includegraphics[width=8.0cm]{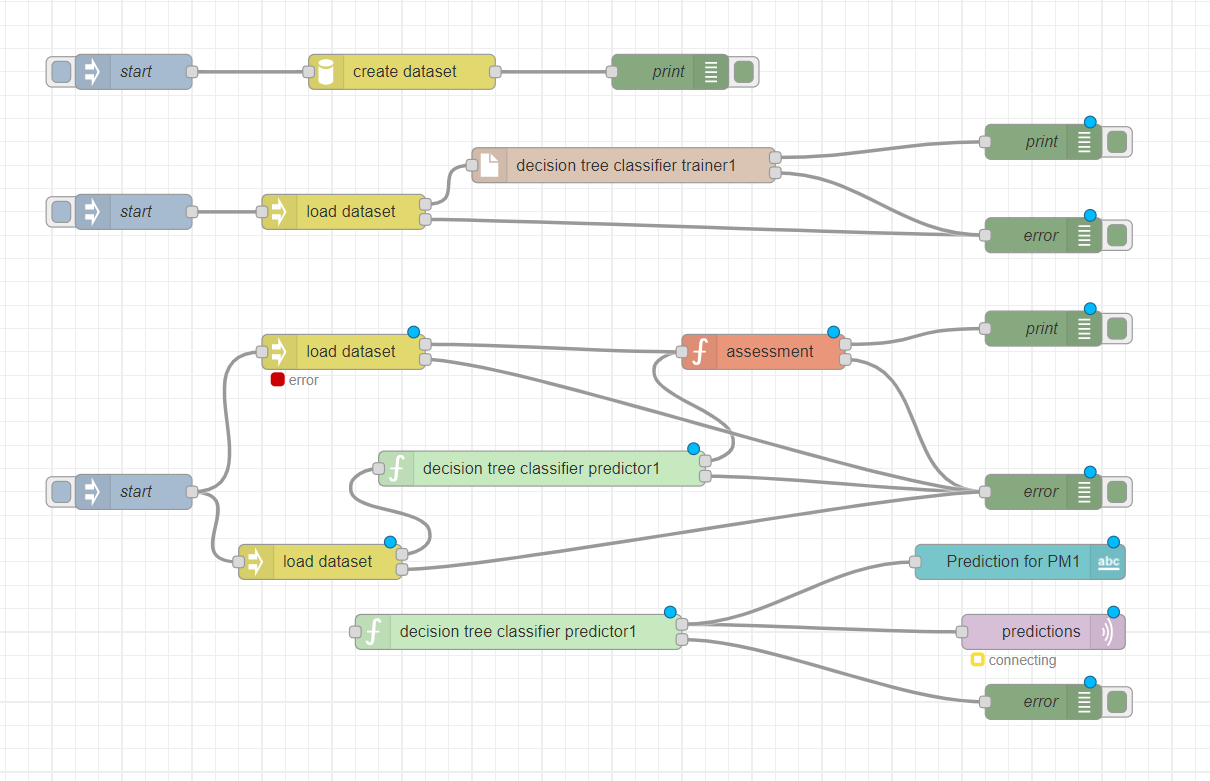}}
\end{minipage}
\caption{Flow management design for deploying the machine learning model with (1) creating and extending dataset with user-requested sensor data (2) training the dataset (3) testing the model for time-series prediction }
\label{fig:res}
\end{figure}

 The publish-subscribe network protocol between NodeRED and NodeMCU is established by Mosquitto broker \cite{Nweb5} in the application layer. The NodeMCU web server is implemented through an Arduino code. Then the processed data is communicated  to the mobile end-user by the mobile application through NodeMCU server and the users can even view the processed data from NodeRED dashboard via the subscription through Mosquitto. 
 
 Here, ESP8266 is used to implement the local server for the data transmission between NodeRED dashboard and mobile or web application.  Data communication between ESP8266 and NodeRED occurs through the MQTT communication protocol via the selected broker service of Mosquitto. Therefore, NodeMCU is working as a subscriber to the NodeRED framework under publish-subscriber architecture, while the NodeRED dashboard works as a client for the primary server sensor network under a server-client architecture. Data communication between ESP8266 and mobile application occurs through the hyper-text transfer protocol (HTTP). A web application is also implemented to communicate with the web server and view the requested data. Since the ESP8266 is utilized as the local server, the mobile application and ESP8266 are connected within the same Wi-Fi network.

The following resources are utilized to implement the NodeMCU server.

\begin{itemize}[noitemsep,nolistsep]
    \item   “ESP8266WiFi.h Arduino library
    \SubItem{To connect with the Wi-Fi network }
    \item   "PubSubClient.h Arduino Library 
    \SubItem{To implement the MQTT broker}
    \item   "ESP8266WebServer.h
    \SubItem{To create a local web server}
\end{itemize}

Finally, the web application and the mobile application are used to access the requested data in all present, past, and forecast formats via real-time notifications. The mobile application is implemented using Android Studio 4.2.2 while facilitating the users to obtain air quality data in the current location. As an additional feature, the location coordinates of the user can be automatically generated and hence, his location can be viewed on a map via the Google Maps application which is embedded in the mobile application. The generated or inserted location coordinates can further be used to request the corresponding air quality data in real-time. 

The following resources are utilized to implement the mobile application.

\begin{itemize}[noitemsep,nolistsep]
    \item   Android Studio \cite{C2}
    \SubItem{Android Studio 4.2.2 version is used as the integrated development environment for the mobile application development. This mobile application is based on a default project in Android Studio which has modalities with source codes and resource files. These modalities include android application modules, library modules, and google application engine modules. Java is used as the programming language in developing the mobile application.}
    \item   Volley HTTP Library \cite{C3}
    \SubItem{Volley, a HTTP library which is capable of building a network for mobile applications is used here. It also provides the facility of automatic scheduling of network requests.} 
    \item  Postman API client \cite{G1}
    \SubItem{The Postman API client allows users to create and save both simple and complex HTTP/s requests while being capable to read their responses. This API client is employed in the application development project to test the APIs.}
    \item  Google Maps \cite{G2}
    \SubItem{Google Maps application is used to display the current location of the user and navigate the direction of the location.}
  
\end{itemize}

Further, following resources are employed to implement the web application.
\begin{itemize}[noitemsep,nolistsep]
    \item   Hyper-text markup language (HTML)
    \SubItem{To design the content to be displayed in the web page}
    \item   Cascaded style sheets (CSS)
    \SubItem{To describe the presentation of the document written HTML}
\end{itemize}

\section{Results}
\label{sec:typestyle}

The proposed system is consisted of both software implementation and hardware implementation. In the hardware implementation, ESP8266 is used to create a local server which is capable of interacting with the web application, NodeRED dashboard, and the mobile application which is integrated with the android operating system.

\subsection{NodeRED Dashboard}
The end-user can enter the coordinates of the location where he requires to obtain the air quality. Then the NodeRED dashboard visualizes the temporal variations and the hourly-based sensor data, which is given by the system, through a wide array of charts. Previous hour parameter values can also be obtained and compared with present and forecast values.

\begin{figure}[htb]

\begin{minipage}[b]{1.0\linewidth}
  \centering
  \centerline{\includegraphics[width=8.5cm]{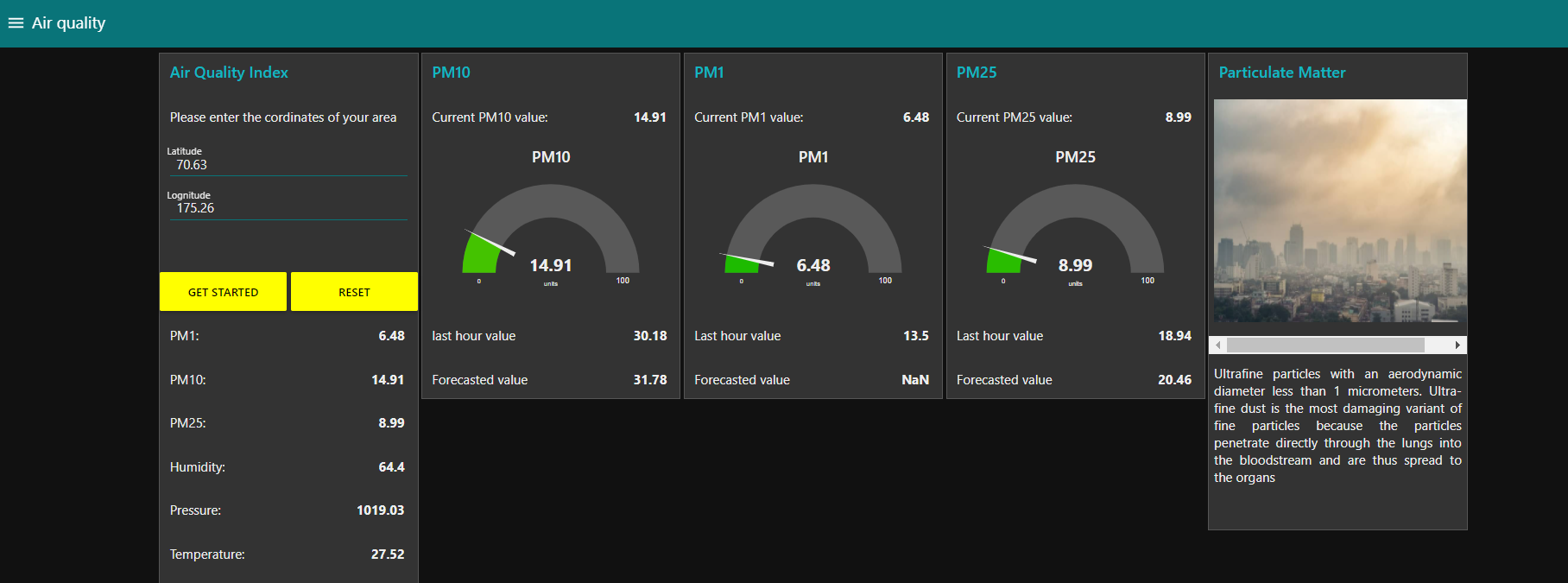}}
\end{minipage}

\caption{The proposed NodeRED dashboard which visualizes the primary and forecast air quality and weather data}
\label{fig:res}
\end{figure}

Furthermore, the end-user can request and visualize the immediate past data and obtain the forecast data through the API and the trained machine learning model. Each air quality parameter has a safe level and if the parameter values of the requested location are exceeding the corresponding safe levels, the system will inform the user about the threat alert through an emergency email.

\begin{figure}[htb]
\begin{minipage}[b]{1.0\linewidth}
  \centering
  \centerline{\includegraphics[width=5cm]{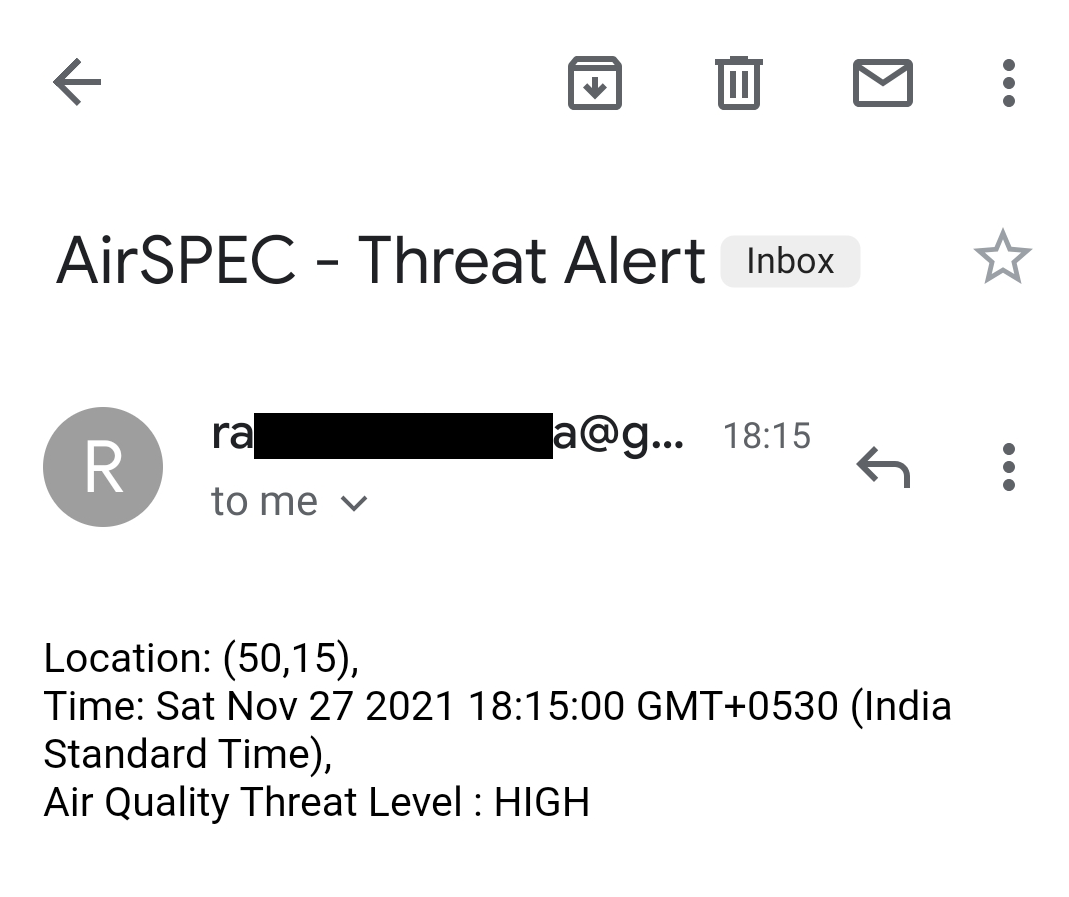}}
\end{minipage}
\caption{The emergency email which is sent by the NodeRED framework for an end-user when the air quality index value of the requested location is above the recommended safe level}
\label{fig:res}
\end{figure}

The dashboard is integrated to work with the MQTT requests through the NodeMCU local server, web and mobile applications. In the NodeRED dashboard, the users who have access to the dashboard, can observe the air quality parameters on the relevant page of the dashboard.

\subsection{Mobile Application}

The user can click the "obtain location coordinates" button on the mobile application interface and enter the latitude and longitude values of the preferred location. According to the requested location, the corresponding results are added to the result tab of the mobile application. In the result tab, the user can observe the sent data through various forms:

\begin{itemize}[noitemsep,nolistsep]

    \item   Current data
    \SubItem {Current sensor data values sent by the system}
    \item   Last 24 hours data
    \SubItem{Sensor data values, which relate to the time before 24 hours at the requested location, stored by the system}
    \item   Forecast data
    \SubItem{Predicted air quality parameter values for three next hours with hourly basis}
\end{itemize}

Users can visualize the required outputs by clicking the "obtain current data" button, "obtain Last 24 hours data" button, and "obtain forecasted data" button. The user interface of the mobile application is simple, attractive, and user-ergonomic. Additionally, mobile users have an option to connect to the Google Maps application as a free add-on.

%%%%%%%%%%%%%%%%%%%%%%%%%%%%%%%%%%%%%%%%%%%%%%%%%%%%%%%
%%%%%%%%%%%%%%%%%%%%%%%%%%%%%%%%%%%%%%%%%%%%%%%%%%%%%%%
\begin{figure}[htb]
\begin{minipage}[b]{.3\linewidth}
  \centering
  \centerline{\includegraphics[width=2.0cm]{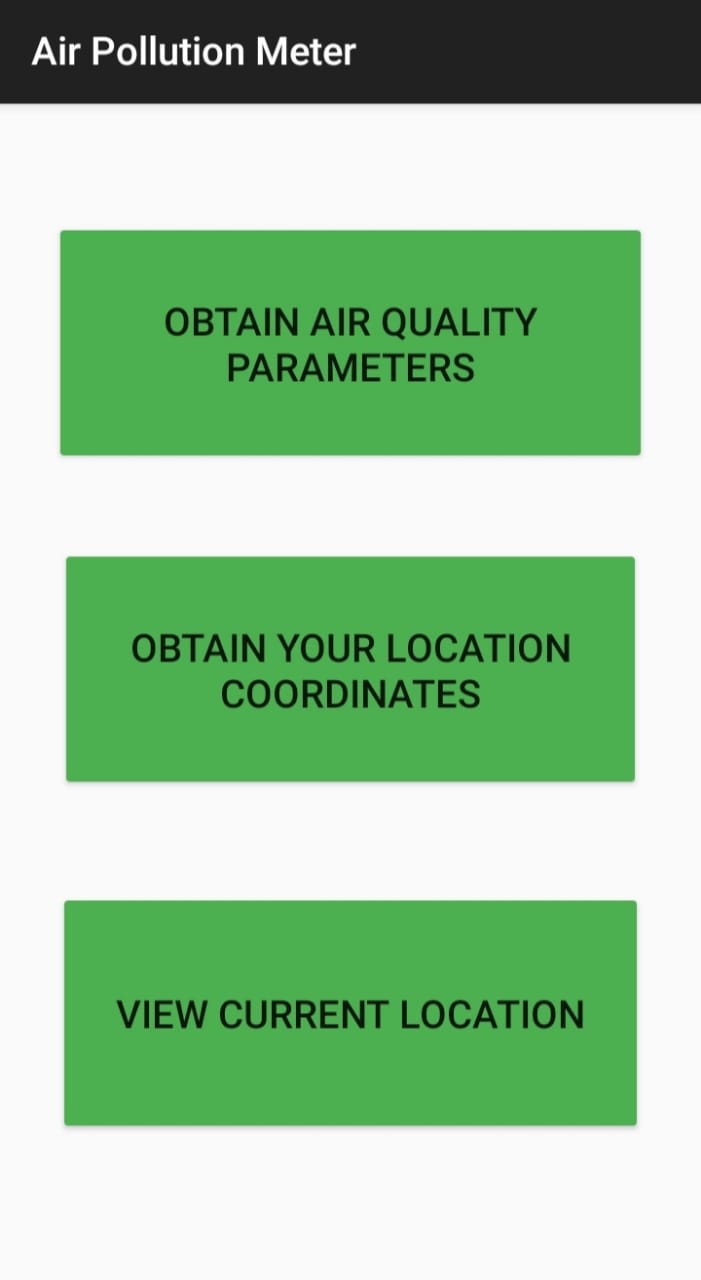}}
%  \vspace{1.5cm}
  %\centerline{(b) Results 3}\medskip
\end{minipage}
\hfill
\begin{minipage}[b]{0.3\linewidth}
  \centering
  \centerline{\includegraphics[width=2.0cm]{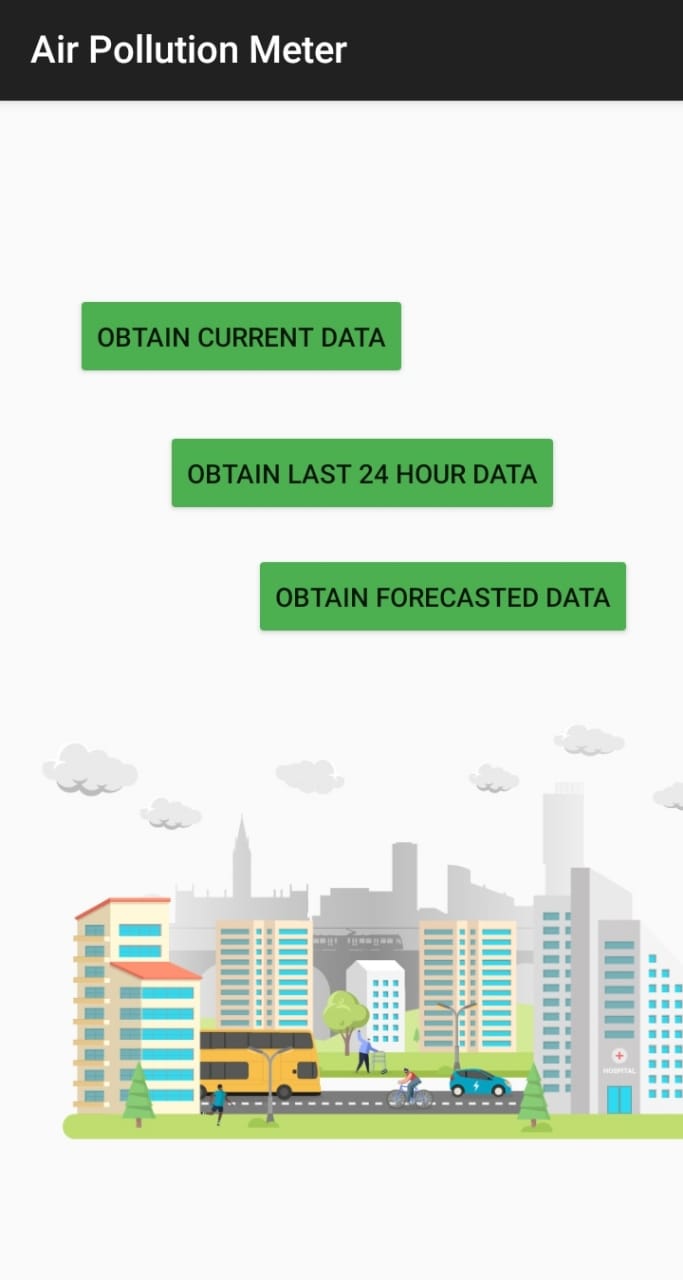}}
%  \vspace{1.5cm}
  %\centerline{(c) Result 4}\medskip
\end{minipage}
\hfill
\begin{minipage}[b]{0.3\linewidth}
  \centering
  \centerline{\includegraphics[width=2.0cm]{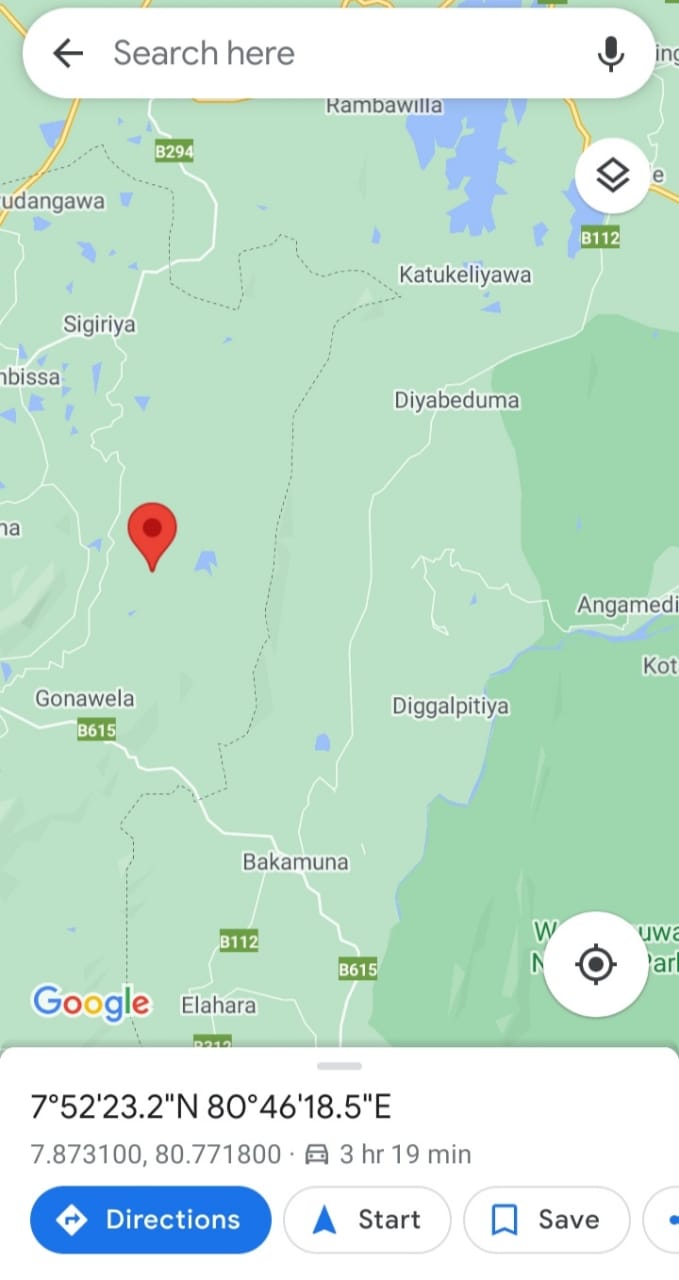}}
%  \vspace{1.5cm}
  %\centerline{(c) Result 4}\medskip
\end{minipage}
\caption{Mobile application interface with three pages for requesting the preferred air quality data and integrating with Google Maps application}
\label{fig:res}
\end{figure}
\vspace{-0.35cm}
\subsection{Web Application}

\begin{figure}[htb]
\begin{minipage}[b]{.3\linewidth}
  \centering
  \centerline{\includegraphics[width=2.0cm]{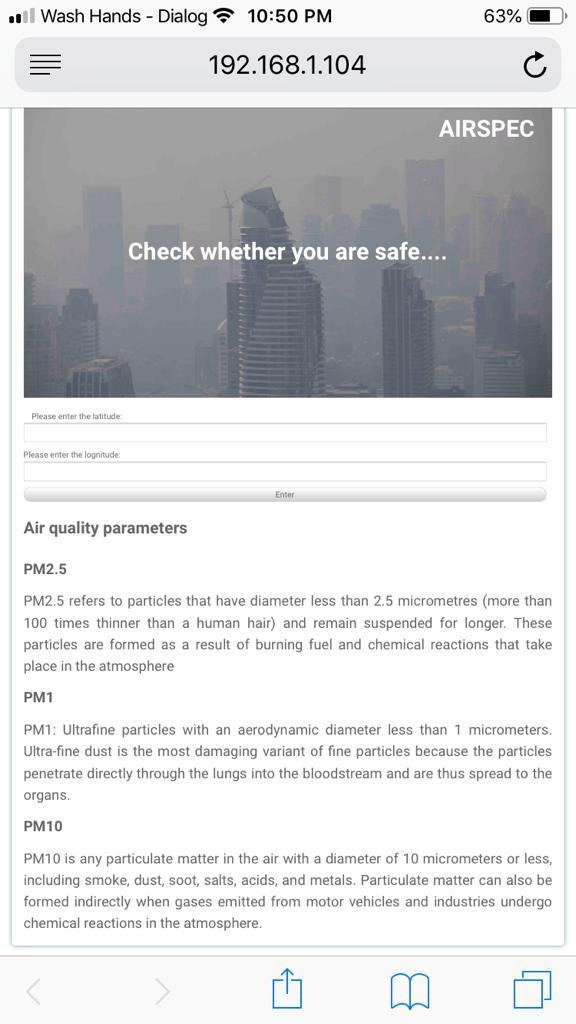}}
%  \vspace{1.5cm}
  %\centerline{(b) Results 3}\medskip
\end{minipage}
\hfill
\begin{minipage}[b]{0.3\linewidth}
  \centering
  \centerline{\includegraphics[width=2.0cm]{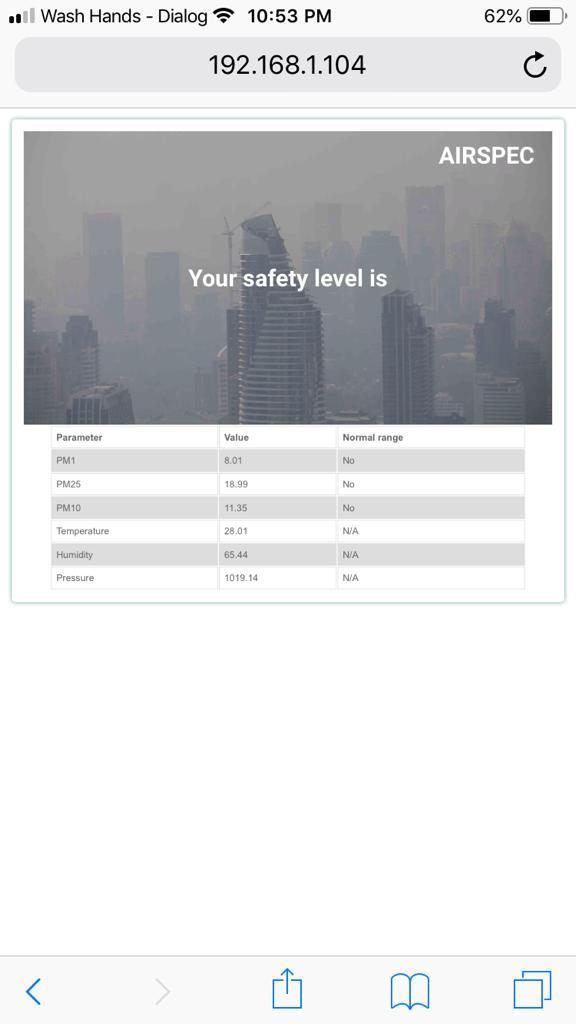}}
%  \vspace{1.5cm}
  %\centerline{(c) Result 4}\medskip
\end{minipage}
\hfill
\begin{minipage}[b]{0.3\linewidth}
  \centering
  \centerline{\includegraphics[width=2.0cm]{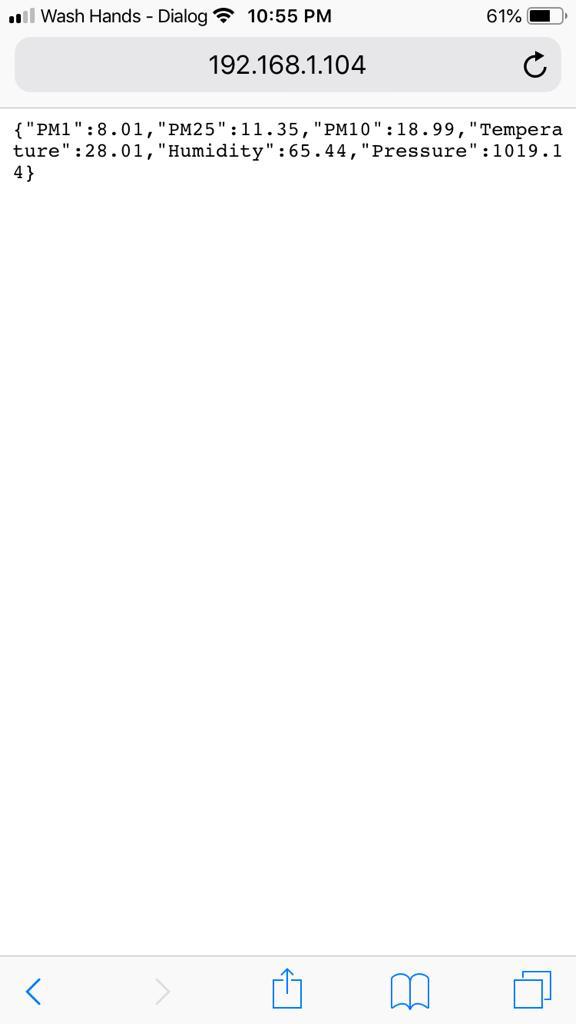}}
%  \vspace{1.5cm}
  %\centerline{(c) Result 4}\medskip
\end{minipage}
\caption{Web application interface with homepage to enter the coordinates of a specific location in order to obtain the air quality data and the corresponding result pages displaying the requested data with respect to recommended range of each air quality parameter}
\label{fig:res}
\end{figure}

Web users can input the preferred location coordinates on the web application's homepage, and the corresponding output will be directed to the results page. The user can obtain a clear idea about the air quality parameters and their standard definitions from WHO \cite{N2} on the homepage. On the result page, web users can observe the corresponding air quality parameter values according to their requested locations and the average safe range of those parameters. 

%%%%%%%%%%%%%%%%%%%%%%%%%%%%%%%%%%%%%%%%%%%%%%%%%%%%%%%

%%%%%%%%%%%%%%%%%%%%%%%%%%%%%%%%%%%%%%%%%%%%%%%%%%%%%%%%
\section{Conclusion}
\label{sec:majhead}

The proposed system presents a novel, semantically distributed, easily expandable, and real-time IoT framework empowered by a machine learning model to identify and forecast air quality parameters in a low-cost implementation. The NodeRED framework obtains primary data from airly and the integrated NodeRED dashboard processes, visualizes, and stores the collected air quality and weather data as a client of the sensor network. End-users may access sensor data as well as forecast data through quantitative and visual representations via built-in mobile and web applications.

Since the primary sensor data is from airly, the proposed system has less capability in controlling primary sensor data. Therefore, in order to extend the capability of the system in acquiring primary sensor data, the system can be integrated with an alternative on-site sensor network to obtain the localized primary sensor air quality data. Furthermore, the proposed system has the potential to be developed as a real-time, accurate, and location-precise health alarming system in future. 
\section{Acknowledgments}
\label{sec:acknowledgments}

Authors would like to extend their gratitude to Prof. Dileeka Dias, the Department of Electronic and Telecommunication Engineering (ENTC), University of Moratuwa for providing valuable guidance. Further, we would like to thank our colleagues at ENTC for their helpful suggestions and feedback.

% References should be produced using the bibtex program from suitable
% BiBTeX files (here: strings, refs, manuals). The IEEEbib.bst bibliography
% style file from IEEE produces unsorted bibliography list.
% ------------------------------------------------------------------------- 
\bibliographystyle{IEEEtran}
\bibliography{strings,refs}

\end{document}